\documentclass{pasj00}

\def\labelspace{}

\def\pasjcite2sub#1#2{\authorcite{#1} \yearcite{#1},\yearcite{#2}}
\def\pasjcitet2sub#1#2{\authorcite{#1} (\yearcite{#1},\yearcite{#2})}
\def\pasjcitep2sub#1#2{(\authorcite{#1} \yearcite{#1},\yearcite{#2})}
\def\pasjcite2subyear#1#2#3{\authorcite{#1} \yearcite{#2}#3}
\def\pasjcitet2subyear#1#2#3{\authorcite{#1} (\yearcite{#2}#3)}
\def\pasjcitep2subyear#1#2#3{(\authorcite{#1} \yearcite{#2}#3)}

\DeclareAbbreviation\an{Astron. Nachr.}
\DeclareAbbreviation\AcA{Acta Astron.}
\DeclareAbbreviation\Ap{Astrophysics}
\DeclareAbbreviation\ARep{Astron. Rep.}
\DeclareAbbreviation\ATsir{Astron. Tsirk.}
\DeclareAbbreviation\BaltA{Baltic Astron.}
\DeclareAbbreviation\ibvs{Inf. Bull. Variable Stars}
\DeclareAbbreviation\JAVSO{J. American Assoc. Variable Star Obs.}
\DeclareAbbreviation\JBAA{J. British Astron. Assoc.}
\DeclareAbbreviation\MitVS{Mitteil. Ver\"{a}nderl. Sterne}
\DeclareAbbreviation\MmSAI{Mem. Soc. Astron. Ita.}
\DeclareAbbreviation\Msngr{Messenger}
\DeclareAbbreviation\NewAR{New Astron. Rev.}
\DeclareAbbreviation\Obs{Observatory}
\DeclareAbbreviation\PAZh{Pis'ma AZh}
\DeclareAbbreviation\PZ{Perem. Zvezdy}
\DeclareAbbreviation\PZP{Perem. Zvezdy Pril.}
\DeclareAbbreviation\RMxAA{Rev. Mexicana Astron. Astrof.}
\DeclareAbbreviation\VeSon{Ver\"{o}ff. Sternw. Sonneberg}
\DeclareAbbreviation\VSOLJBul{VSOLJ Variable Star Bull.}
\DeclareAbbreviation\ZA{Z. Astrophys.}

\begin{document}
\SetRunningHead{M. Uemura et al.}{A Dwarf Nova breaking the Standard
Evolutionary Scenario}
\Received{2001/12/11}
\Accepted{2002/01/21}

\title{Discovery of a Dwarf Nova Breaking the Standard Sequence of
Compact Binary Evolution}

\author{
Makoto \textsc{Uemura},\altaffilmark{1}
Taichi \textsc{Kato},\altaffilmark{1}
Ryoko \textsc{Ishioka},\altaffilmark{1}
Hitoshi \textsc{Yamaoka},\altaffilmark{2}
Patrick \textsc{Schmeer},\altaffilmark{3}\\
Donn R. \textsc{Starkey},\altaffilmark{4}
Ken'ichi \textsc{Torii},\altaffilmark{5}
Nobuyuki \textsc{Kawai},\altaffilmark{5}\altaffilmark{6}
Yuji \textsc{Urata},\altaffilmark{5}\altaffilmark{7}
Mitsuhiro \textsc{Kohama},\altaffilmark{5}\\
Atsumasa \textsc{Yoshida},\altaffilmark{5}\altaffilmark{8}
Kazuya \textsc{Ayani},\altaffilmark{9}
Tetsuya \textsc{Kawabata},\altaffilmark{9}
Kenji \textsc{Tanabe},\altaffilmark{10}\\
Katsura \textsc{Matsumoto},\altaffilmark{11}
Seiichiro \textsc{Kiyota},\altaffilmark{12}
Jochen \textsc{Pietz},\altaffilmark{13}
Tonny \textsc{Vanmunster},\altaffilmark{14}\\
Tom \textsc{Krajci},\altaffilmark{15}
Arto \textsc{Oksanen},\altaffilmark{16} and 
Antonio \textsc{Giambersio}\altaffilmark{17}
}

\altaffiltext{1}{Department of Astronomy, Faculty of Science, Kyoto University,
  Sakyou-ku, Kyoto 606-8502}
\email{uemura@kusastro.kyoto-u.ac.jp}
\altaffiltext{2}{Faculty of Science, Kyushu University, Fukuoka 810-8560}
\altaffiltext{3}{Bischmisheim, Am Probstbaum 10, 66132 Saarbr\"{u}cken, 
Germany}
\altaffiltext{4}{AAVSO, 2507 County Road 60, Auburn,, Auburn, Indiana 46706, USA}
\altaffiltext{5}{Cosmic Radiation Laboratory, Institute of Physical and
Chemical Research (RIKEN), 2-1, Wako, \\Saitama 351-0198}
\altaffiltext{6}{Department of Physics, Faculty of Science, Tokyo Institute
of Technology 2-12-1 Ookayama, \\Meguro-ku, Tokyo 152-8551}
\altaffiltext{7}{Department of Physics, The Science University of Tokyo,
Kagurazaka, Shinjuku-ku, \\Tokyo 162-8601}
\altaffiltext{8}{Department of Physics, Aoyama Gakuin University, 6-16-1,
Chitosedai, Setagaya-ku, \\Tokyo 157-8572}
\altaffiltext{9}{Bisei Astronomical Observatory, Ohkura, Bisei, Okayama
714-1411}
\altaffiltext{10}{Department of Biosphere-Geosphere Systems, Faculty of 
Informatics, \\Okayama University of Science, Ridaicho 1-1, Okayama
700-0005}
\altaffiltext{11}{Graduate School of Natural Science and Technology, Okayama
University, Okayama 700-8530}
\altaffiltext{12}{Variable Star Observers League in Japan (VSOLJ);
  Center for Balcony Astrophysics, \\1-401-810 Azuma, Tsukuba 305-0031}
\altaffiltext{13}{Nollenweg 6, 65510 Idstein, Germany}
\altaffiltext{14}{Center for Backyard Astrophysics (Belgium), Walhostraat 1A, B-3401, Landen, Belgium}
\altaffiltext{15}{1688 Cross Bow Circle, Clovis, New Mexico 88101, USA}
\altaffiltext{16}{Nyrola Observatory, Jyvaskylan Sirius ry,
Kyllikinkatu 1, FIN-40100 Jyvaskyla, Finland}
\altaffiltext{17}{AAVSO Via Rocco Scotellaro, 22 85100 Potenza, Italy}

%

\KeyWords{accretion, accretion disks---stars: binaries: close---individual
(1RXS J232953.9+062814)} 

\maketitle

\begin{abstract}
 We revealed that the dwarf nova 1RXS J232953.9+062814 is an SU UMa-type
 system with a superhump period of $66.774\pm 0.010\; {\rm min}$.  The 
 short period strongly indicates that the orbital period of this object
 is below the period minimum of cataclysmic variables.  The superhump
 period is $4.04\pm 0.02$\% longer than the photometric period during 
quiescence 
 ($64.184\pm 0.003$ min), which is probably associated with the orbital
 period.  Although the standard evolutionary scenario of cataclysmic
 variables predicts lower mass-transfer rates in systems with shorter
 orbital periods, we revealed firm evidence of a relatively high
 mass-transfer rate from its large proper motion and bright apparent
 magnitude.  Its proximity indicates that we have overlooked a number of
 objects in this new class.  With the analogous system of V485 Cen,
 these objects establish the first subpopulation in hydrogen-rich
 cataclysmic variables below the period minimum.
\end{abstract}

\section{Introduction}
Cataclysmic variables (CVs) are compact binary systems in which the surface
gas of a secondary star overflows and is accreted by a more massive 
white dwarf (\cite{war95book}).  The long-lived mass-transfer is maintained 
by the continuous removal of their orbital angular momentum.  In systems 
with short ($\lesssim 2\;{\rm hr}$) orbital periods, it is believed that 
the mass-transfer rate is governed by the angular-momentum loss caused 
by gravitational radiation (\cite{taa80GWR}).  Losing angular 
momentum, systems evolve into those with shorter orbital periods, 
smaller secondaries, and hence lower mass-transfer rates.  When the 
secondary star becomes degenerate, a decrease of mass leads to an expansion 
of the secondary, and then, an increase of the orbital period.  The above 
scenario has been applied to explain the observed ``period minimum'' of
about 80 min, and has been widely accepted as the standard evolution
model of compact binary systems (\cite{pac81CVevolution};
\cite{kin88CVevolution}).   
 
1RXS J232953.9+062814 was discovered as an X-ray source with the 
R\"ontgen Satellite (ROSAT) X-ray telescope (\cite{vog96rosat}).  
The object is identified with an optical source whose $V$-magnitude is
15.7 (\cite{jin98J2329}).  Optical spectroscopy revealed two distinct
states of this object (\cite{jin98J2329}).  One is a faint state in
which the optical spectrum is dominated by hydrogen emission lines,
indicating that this object is a hydrogen-rich CV.  Noteworthy features 
during this state are the relatively strong He~{\sc i} emission and TiO 
absorption bands, the latter being typical for M-type stars.  The other 
state is a bright one in which the hydrogen lines appear in
absorption.  Based on these observations, this object has been 
classified as a dwarf nova (DN), a sub-group of CVs which experience 
repetitive outbursts with typical amplitudes of 2--5 mag 
(\cite{war87DNreview}).  DN outbursts are considered to be a suddenly 
enhanced release of gravitational energy induced by the thermal
instability of an accretion disc (\cite{osa96review}).  

Here, we report on the first detailed photometric observations of this
object, including an outburst detected in 2001 November 3, in which we 
reveal that this object is an ultrashort period system with a relatively 
high mass-transfer rate.  Our detailed results concerning superhump 
evolution in this system will be reported in a forthcoming paper. 
  
\section{Observation}
Our CCD photometric observations were performed with 30-cm class
telescopes from 2001 November 4 to December 4 at Kyoto (26 nights), 
Auburn (9), Wako (3), Bisei (1), Okayama (5), Tsukuba (5), Idstein 
(3), Landen (3), Clovis (10), Nyrola (2), and Potenza (1).  The
exposure time was 30--120 s.  After dark subtraction and flat 
fielding on the images, we performed aperture photometry and obtained 
differential magnitudes of the object using a comparison star, 
GSC 591.1689.  The constancy of the comparison star was checked by 
GSC 584.366.  The magnitude scales of each observatory were adjusted to 
that of the Kyoto system.  We could obtain magnitudes almost equal to 
the $R_{\rm c}$ system from observations at Kyoto in which we used an 
unfiltered 
ST-7E CCD camera, since the sensitivity peak of the camera is near that
of the $R_{\rm c}$ system and the color of the object is $B-V\sim 0$.  
Heliocentric corrections to the observed times were applied before the 
following analysis.

\section{Result}
\subsection{Superoutburst in 2001 November}

Followig our detection of an outburst of this object on 2001 November 3 at 
12.5 mag, we started CCD optical monitoring.  Figure 1a shows the whole 
light curve during this outburst.  The object gradually faded at a
rate of $0.25\;{\rm mag\, d^{-1}}$ for the first 5 days.  This decline 
rate is much slower than those observed in ordinary outbursts of DNe, 
and rather similar to those in superoutbursts of SU UMa stars.   
This plateau phase lasted for at least four days, and then the object
rapidly faded.  After the main outburst, we detected a short
rebrightening starting on 2001 November 9, as can be seen in figure 1a.  
The brightness then gradually declined for about one week and returned to
the quiescent level around 2001 November 18. 

\begin{figure}
  \begin{center}
    \FigureFile(80mm,80mm){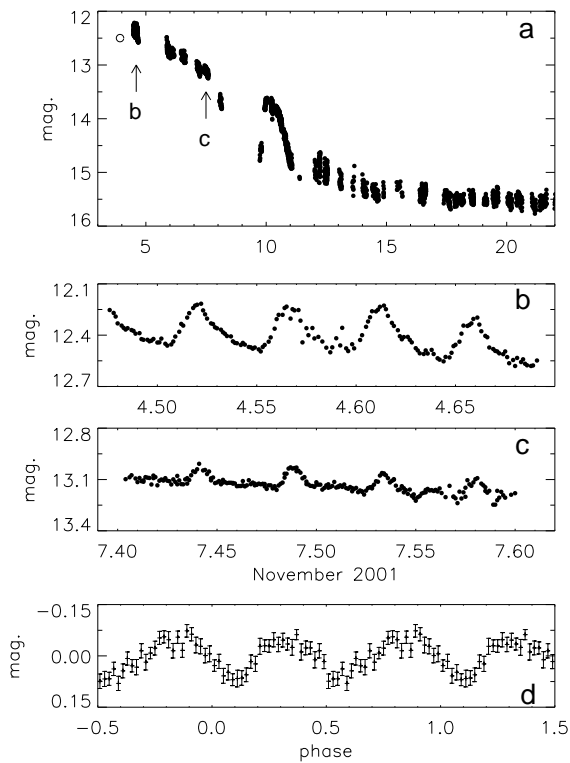}
  \end{center}
  \caption{Light curve of 1RXS J232953.9+062814 during the
 outburst on 2001 November.  (a): The whole light curve of the outburst.
 The abscissa and ordinate denote the date and $R_{\rm c}$ magnitude, 
 respectively.  The open circle denotes the discovery of this outburst
 with the visual estimation performed by one of the authors (P.S.).  (b)
 and (c): Light curves on November 4 and 7, respectively.  The
 corresponding dates in these panels are marked in panel a.  (d):
 Average light curve on November 19 -- December 4.  We folded the light curves 
 with the period of 64.184 min.  The abscissa and ordinate denote the 
 phase in this period and the relative magnitude, respectively.  The
 epoch of the phase is arbitrary.}
\label{fig:lc}
\end{figure}

During the plateau phase, we discovered periodic modulations of 
brightness.  The modulations appeared throughout the outburst, and even
in the rebrightening and early quiescent phase.  We show their typical 
light curves in figures 1b and c.  The humps have a common profile of
rapid-rise and gradual-decline, while their amplitudes decreased with 
time from about 0.25 to 0.10 mag.  After we subtracted the linear fading 
trend from the light curve on November 4--7, we performed a period
analysis on the humps during the plateau phase using the Phase
Dispersion Minimization (PDM) method (\cite{PDM}).  The best candidate
of the period was then calculated to be $66.774\pm 0.010$ min. 

The characteristics of the humps are the same as those of superhumps,  
which are periodic modulations commonly appearing in SU UMa stars during 
their superoutbursts (\cite{odo00SH}).  The period of superhumps are 
generally a few percent longer than the orbital period, which is now 
interpreted as a beat phenomenon between the orbital period and the 
period of slow apsidal motion of an elongated accretion disc caused by 
a tidal torque from the secondary star (\cite{whi88tidal}; 
\cite{osa89suuma}).  Using the light curve in the quiescent phase 
(November 19 -- December 4), our PDM period analysis yielded a period of 
$64.210\pm 0.023$ min.  The period of the humps during the outburst was 
$4.036\pm 0.015$\% longer than this quiescent period.  As shown in
figure 1d, the profile of modulations during the quiescent phase was a
double-peaked sinusoidal curve, which is completely different from those
during the outburst, and hence, indicates their different natures.  In
conjunction with these two periodicities, we conclude that the periods
during the outburst and quiescence can be identified by the superhump
and orbital period, respectively.  Using observations before this
outburst, Zharikov and Tovmassian (2001)
\footnote{$\langle$http://www.kusastro.kyoto-u.ac.jp/vsnet/Mail/vsnet-alert/msg06851.html$\rangle$} 
reported a period of $64.2\pm 0.1\;{\rm min}$, which is in complete 
agreement with our estimated one.  This strongly supports the constancy 
of this period during 
quiescence, and that it is the orbital period.   
1RXS J232953.9+062814 hence breaks the observed period minimum of about 
80 min, which appears in hydrogen-rich CVs, as shown in figure 2.  

\subsection{Estimation of the distance and absolute magnitude}

The optical flux of a CV is dominated by emission from an accretion disc, 
particularly in short-period
systems which have a very low-luminosity secondary.  Since the emission 
from the disc is proportional to the mass-transfer rate, the absolute 
magnitudes ($M_V$) of CVs are good indicators of 
the mass-transfer rates (\cite{spr96CVabsmag}).  To determine the absolute 
magnitude of an object, we need its distance.  We applied two
independent methods to set a limit on the distance of 1RXS
J232953.9+062814, i.e., the transverse motion on the celestial plane and
the famous empirical relation between the peak brightness and the
orbital period ($P_{\rm orb}$) (\cite{war95book}).  

Using our images taken on 2001 November, we measured the position of the 
object to be R.A.$=23^{\rm h}29^{\rm m}54^{\rm s}.23$ and 
Dec. $=+06^\circ 28^\prime12^{\prime \prime}.4$ using template stars
in the USNO-A2.0 catalogue (\cite{mon98usno}).  On the other hand, we
obtained the position in 1951.6 using the USNO-A2.0 catalogue, in which it
is reported to be R.A. $=23^{\rm h}29^{\rm m}54^{\rm s}.357$ and 
Dec.$=+06^\circ 28^\prime 10^{\prime \prime}.35$.  From these positions, 
we obtained significant proper motions of 
$\Delta {\rm R.A.} = 38\; {\rm mas\,yr^{-1}}$ and 
$\Delta {\rm Dec.} = 41\; {\rm mas\, yr^{-1}}$.  
To calculate a secure upper-limit of the distance, we neglected the radial 
velocity and considered the transverse velocity of this object to be 
$100\;{\rm km\, s^{-1}}$, which corresponds to the maximum expected 
velocity dispersion of CVs (\cite{har00DNdistance}).  
With the above proper motion, it yields a distance of $< 380\; {\rm
pc}$.  

With the framework of the disc-instability model for DN outbursts, the 
peak luminosities strongly depend on the amount of mass stored in
the disc, namely the size of the disc and the binary system 
(\cite{osa96review}).  The well-known empirical law  
$M_V({\rm peak})=5.74-0.259P_{\rm orb}({\rm hr})$ can be interpreted
with this theoretical prediction (\cite{war95book}).  The peak magnitude
in this equation is not those in superoutbursts which are accompanied
by superhumps, but in normal ones whose peak magnitudes are generally 
fainter than those of superoutbursts.  We can thus obtain a lower limit 
of the distance from our observations of the superoutburst of 1RXS
J232953.9+062814.  The above equation indicates that the 64.2-min
orbital period yields a peak absolute magnitude of 5.46 mag.  With the  
observed apparent magnitude of 12.41 mag, we calculated the lower-limit
of the distance to be $\sim 245 \;{\rm pc}$, while neglecting intersteller 
extinction and a term concerning the inclination of the disc.    

\begin{figure}
  \begin{center}
    \FigureFile(80mm,80mm){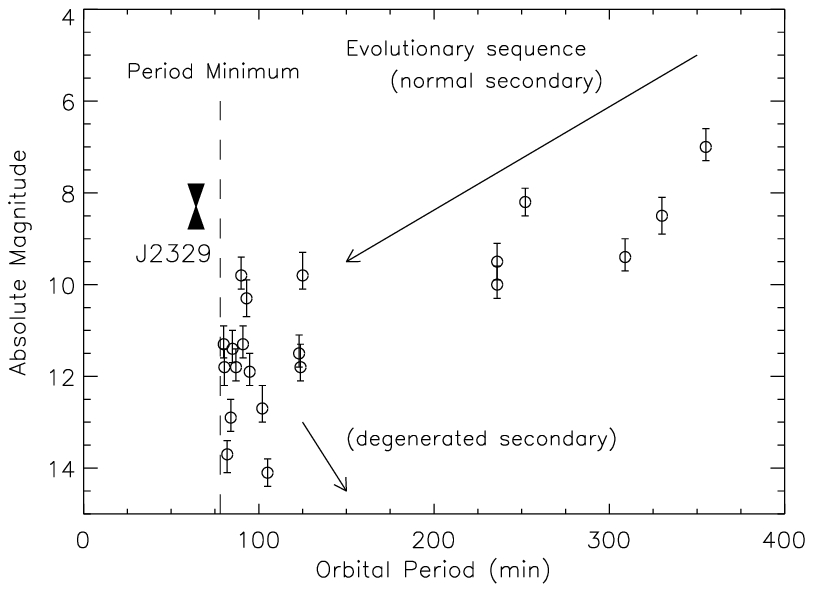}
  \end{center}
  \caption{Absolute magnitude of CVs as a function
 of their orbital period.  We show the lower and upper limit of the
 quiescent brightness of 1RXS J232953.9+062814 as filled triangles and
 indicated by 'J2329'.  The open circles denote DNe listed in
 \citet{spr96CVabsmag}.  The arrows are schematic evolutionary sequences
 expected from theoretical models in which the orbital angular momentum
 is continuously removed from systems due to gravitational
 radiation.  The upper and lower arrows correspond to that of
 CVs with a normal secondary star and a degenerated
 one, respectively.  The vertical dotted line shows the observed period
 minimum (80 min).}
\label{fig:absmag}
\end{figure}

The above two estimations of the distance are consistent with each other and
yield a quiescent absolute magnitude of $+7.8 < M_V < +8.8$ using a  
quiescent $V$-magnitude of 15.7.   The estimated quiescent brightness is 
surprisingly high.  DNe at quiescence are generally much fainter 
than this object around the orbital period just above the period minimum
($M_V=10$--$14$, see figure 2) (\cite{spr96CVabsmag}).  We can
estimate the absolute magnitude of the secondary star to be fainter than 
12.7 mag under the condition of a secondary star filling the Roche lobe
of a system with $P_{\rm orb}=64.2\; {\rm min}$ and having the effective 
temperature of an M-type main-sequence (\cite{jin98J2329};
\cite{wei01J2329}).  This faint 
secondary star cannot explain the observed absolute magnitude.  We thus 
consider that the optical flux is dominated by the accretion-disc 
emission in 1RXS J232953.9+062814, as in other short-period systems.  
This is also supported by the quiescent optical spectrum, since it shows 
the 
continuum much bluer than that in a single M-type star (\cite{jin98J2329}).  
We therefore conclude that it has an intrinsically high mass-transfer
rate, which completely contradicts the standard evolutionary
scenario, as depicted in figure 2.  

\section{Discussion and Summary}
CVs which have a secondary star of a helium white
dwarf are known to have orbital periods shorter 
than 1RXS J232953.9+062814 and relatively high mass-transfer rates 
(\cite{ull94amcvn}).  These systems with hydrogen-deficient secondary
stars have been proposed to have evolutionary tracks distinct from
those of ordinary systems with hydrogen-rich stars 
(\cite{sar96}).  On the other hand, the strong hydrogen 
emission lines seen in the quiescent optical spectrum indicate that 
this object does not belong to this class. 

Besides the high mass-transfer rate, another notable feature is the
large superhump excess of $4.04$\%.  Both theoretically and 
empirically,
systems with a smaller mass ratio ($q=M_2/M_1$) are generally expected to 
have smaller superhump excesses (\cite{pat01SH}).  We can understand
this by the weak tidal effect from quite low-mass secondary stars in
short-period systems.  This superhump excess is
ones of the largest one in SU UMa stars, and thus indicates the unexpectedly
large mass ratio of this system.  The empirical relation yields
$q=0.19\pm 0.02$
(\cite{pat01SH}).  On the other hand, although the quiescent optical 
spectrum indicates the presence of a secondary star with an effective 
temperature similar to that of M-type stars, it is certainly too large for the
secondary star of this ultrashort period object (\cite{jin98J2329};
\cite{wei01J2329}).  These arguments imply the presence of a
relatively massive secondary.  In conjunction with the ultrashort
orbital period, it may partly cause the high mass-transfer rate driven 
by the gravitational radiation.  In this case, the secondary star probably 
evolves off the main-sequence, which means a distinct evolutionary
sequence compared with the ordinary hydrogen-rich CVs.  It is possible
that a quite low-mass white dwarf causes the small mass-ratio and a
secondary when a moderate mass star is heated by the UV--X-ray flux from
the accretion disc.  In this case, we expect a temperature
inversion in the atmosphere of the secondary star and the formation of 
emission lines.  The quiescent spectrum, however, shows no evidence for
such lines (\cite{jin98J2329}; \cite{wei01J2329}).

In known hydrogen-rich CVs, we can find one analogous 
object, that is, the DN V485 Cen, whose quiescent apparent 
magnitude is $V=18.4$ and orbital period is 59 min (\cite{aug96v485cen}; 
\cite{aug93v485cen}).  V485 Cen and 1RXS J232953.9+062814 have some 
noteworthy common features, that is, a short duration of outbursts, 
a rebrightening phenomenon, and relatively strong He~{\sc i} emission 
(\cite{ole97v485cen}).  We thus propose that these two objects establish 
a new sub-class below the period minimum.  The short distance of 
1RXS J232953.9+062814 
strongly indicates that we have overlooked a number of objects which 
belong to this class.  

We revealed that 1RXS J232953.9+062814 is an SU UMa-type DN
below the period minimum with a high mass-transfer rate.  The
evolutionary status and the driving mechanism of the angular momentum 
removal of this class make a new issue concerning the late evolution of compact
binaries.  
Theoretical calculations imply that the period minimum depends on the opacities
and hydrogen fraction in the secondary of CVs (\cite{sie84};
\cite{nel86ultrashortPbinary}).  A system with a moderately
hydrogen-deficient secondary is predicted to have a shorter period
minimum and a larger mass-transfer rate before reaching its period 
minimum.  It is thus possible that such a secondary star causes the atypical
system parameters of 1RXS J232953.9+062814.  Only with equivalent 
widths of hydrogen and helium estimated from optical spectra,
however, it is difficult to determine their fraction in the secondary.  
Our discovery of this object with a bright apparent magnitude will
provide a unique chance for us to perform detailed observations, 
including a determination of the primary and secondary masses, and thereby to 
study the evolutionary status of this class, which has been difficult 
with only the faint source V485 Cen.   

\vskip 3mm

We are pleased to acknowledge comments by D. Nogami, H. Baba, and
J. Hu, and vital observations via VSNET by many amateur observers.  
Part of this work is supported by a Research Fellowship of the Japan
Society for the Promotion of Science for Young Scientists (MU).


\end{document}